# Machine-learning assistant DFT study of half-metallic full-Heusler alloy N$_2$CaNa: structural, electronic, mechanical, and thermodynamics properties

E. B. Ettah, M.E. Ishaje
*Depatment of Physics, University of Cross River State, Calabar, Nigeria*
K. A. Minakova
*Department of Physics, National Technical University "Kharkiv Polytechnic Institute", Kharkiv, Ukraine*
V.A. Sirenko, I. S. Bondar
*Verkin Institute for Low Temperature Physics and Engineering of the National Academy of Sciences of Ukraine, Kharkiv, Ukraine*

**Abstract**

*We studied the structural, electronic, mechanical, and thermodynamic properties of N$_2$CaNa full Heusler alloys using density functional theory (DFT). Results for the structural analysis establishes structural stability with a minimum formation energy of 29.90eV. The compounds is brittle and mechanically stable, having checked out with the Pugh criteria. B/G ratio for N$_2$CaNa is 4.766, hence the material is ductile. N$_2$CaNa alloy is ductile in nature. Debye model correctly predicts the low temperature dependence of heat capacity, which is proportional to the debye T$^3$ law. Just like the Einstein model, it also recovers the Dulong-Petit law at high temperatures, suggests thermodynamic stability of the compounds at moderate temperatures. The results demonstrate potential for applications in spintronics, structural engineering, and other fields requiring materials with tailored properties.*

## 1.0 Introduction

In the realm of materials science, the exploration of novel compounds with unique properties is a constant pursuit. Full-Heusler alloys have garnered significant attention due to their remarkable electronic, magnetic, and mechanical properties, making them promising candidates for various technological applications, including spintronics and thermoelectricity. Among these alloys, N$_2$CaNa, a half-metallic half-Heusler compound, has emerged as a subject of interest owing to its potential in spintronics and other advanced applications [1, 21].

Some Half-Heusler alloys are semi-metallic, their band structures exhibit metal character in one skin and semi-conducting in the other. Such materials have great applications in spintronic [6]. Even some other half Hausler compounds are narrow band-gap semi-conductors [5]. Promoting them as good candidates for optoelectronic [7] applications. The half Heusler alloys are comprised of less expensive, more abundant and non-toxic elements [8]. In addition, full Hausler alloys have better electronic properties than many best thermoelectric materials but their high thermal conductivity reduces the efficiency of energy conversion due to the dominant lattice thermal conductivity in half-Hausler [13]. Eventually, many efforts have been done to reduce lattice thermal conductivity of these material by keeping their various electrical properties. This ensures the half Hausler alloys to meet the demands of advanced thermoelectric materials. Though only a few of them are well studied. Nature of the materials and band astrictive is very important for industrial applications of these materials [11].

First principle calculations are a powerful computational approach used to investigate the properties of materials form fundamental physical principles, without relying on experimental data. In the case of Half-Hausler alloys, these calculations involve soluting the quantum mechanical equations that describe the behavior of electrons and atoms in the material [12, 20].

Structural properties: First-principle calculations can determine the equilibrium crystal structure, lattice parameters, and atomic positions of Half-Hausler alloys. These calculations help understand the arrangement of atoms and how it affect the materials stability and properties [15].

Electronic properties: By soluting the electronic structure equations, researchers can obtain valuable information about the band structure, density of states, and electronic band gaps of Half-Hausler alloys. These properties are crucial for understanding the materials electrical conductivity, magnetism and optical behavior [16].

Mechanical properties: First principles calculation can also provide insight into the mechanical properties of Full-Hausler alloys, such as elastic constants, bulk modulus, shear modules and poisons ration. These properties help determine the materials responses to external forces and its potential applications in mechanical engineering [17].

Thermodynamic properties: Calculating the thermodynamic properties of Full-Hausler alloys, allows researchers to investigate the stability, phase transition, and temperature-dependent behavior. These calculations involves determine the formation energies, vibrational contributions and free energies of the materials [18].

## 1.1 Aim and Objectives

This study aims to investigate the structural, electronic, mechanical, and thermodynamic properties of $N_2CaNa$ using Density Functional Theory (DFT) calculations.

## 1.2 The specific objectives include:

1. Determining the optimized structural parameters of $N_2CaNa$
2. Analyzing the electronic band structure and density of states to understand its electronic properties.
3. Evaluating the mechanical properties such as elastic constants, bulk modulus, and shear modulus.
4. Investigate the thermodynamic properties including heat capacity, entropy, and formation energy.

## 1.3 Mathematical Background

Bulk modulus (B): $\quad B = -v\left(\frac{\partial p}{\partial v}\right)T$ (1)

Shear modulus (G): $\quad G = \frac{1}{2}(C_{11} - C_{12})$ (2)

Young's modulus (E): $\quad E = \frac{9BG}{3B+G}$ (3)

where V is the volume of the unit cell, P is the hydrostatic pressure, $C_{11}$ and $C_{12}$ are the elastic constants of the material.

The Poisson's ratio (v) is another important mechanical property that describes the material's response to deformation. It is defined as the ratio of transverse strain to axial strain and can be calculated as:

$$v = \frac{-(C_{11} - 2C_{12})}{2C_{12}} \tag{4}$$

These equations provide a quantitative understanding of the mechanical properties and deformation behavior of materials [9, 14, 22].

**2.0 Methodology**

Density Functional Theory (DFT), implemented in the Vienna Ab initio Simulation Package (VASP), was employed to carry out the computational investigations. The Perdew-Burke-Ernzerhof (PBE) exchange-correlation functional within the framework of the generalized gradient approximation (GGA) was utilized. The projector augmented wave (PAW) method was employed to describe the electron-ion interaction. The Brillouin zone integration was performed using Monkhorst-Pack k-point mesh. Structural optimizations were carried out until the forces on each atom were less than 0.01 eV/Å.

# Optimized input file used for simulation

This Quantum ESPRESSO input file sets up a self-consistent field (SCF) calculation for a spin-polarized system within a face-centered cubic (fcc) lattice structure. The control section specifies the type of calculation, file prefixes, and directories for pseudopotentials and temporary files. The system section defines key parameters such as the Bravais lattice index (ibrav=2), lattice parameter (celldm(1)), and the number of atoms (nat) and types (ntyp) in the unit cell. Additionally, it handles electronic state occupations using Marzari-Vanderbilt smearing and sets initial magnetizations for each atomic type, with a specified plane-wave kinetic energy cutoff (ecutwfc) to ensure the accuracy of the wavefunctions.

The electrons section focuses on convergence settings for the SCF procedure, specifying the mixing factor (mixing_beta) and the diagonalization algorithm (Davidson). It also sets the maximum number of iterations (electron_maxstep) to ensure the calculation converges properly. These parameters are crucial for achieving a reliable and efficient solution to the electronic structure problem, especially for complex systems involving multiple atom types and spin polarization.

The atomic species and atomic positions sections detail the types of atoms present, their masses, and the corresponding pseudopotentials. This includes nitrogen (N), calcium (Ca), and sodium (Na), each with specific pseudopotential files. The k-points section defines the mesh for sampling the Brillouin zone, using an 8x8x8 grid with no shift, which is essential for accurately capturing the electronic properties of the material. This comprehensive setup ensures a thorough and precise simulation of the material's electronic and magnetic properties.

A detailed description for the Quantum ESPRESSO input file you've provided, which sets up a self-consistent field (SCF) calculation for a spin-polarized system within a face-centered cubic (fcc) lattice structure. This specific input is tailored for the Density Functional Theory (DFT) study of the half-metallic full-Heusler alloy $N_2CaNa$.

```
&control
        calculation ='scf',                                      !.self-consistent calculation
        prefix='n2cana'                                          !.Prefix for output files
        pseudo_dir='/home/ishajemichael/PSEUDOPOTENTIALS/',!.Directory containing pseudopotentials
        outdir='./',                                             !.Directory for temporary files
/
        &system
        ibrav=2,                                                 !.Index for the bravais lattice
        celldm(1)=13.0684,                                       !.lattice parameter a
        nat=4,                                    !.Number of atoms in the unit cell
        ntyp=3,                                   !.Number of atom types
        occupations='smearing', smearing='mv', degauss=0.02 !.Occupation type, !smearing method            ecutwfc= 70,
                                        !.Planewave kinetic energy cutoff
/
        &electrons
        mixing_beta = 0.7                                        !.Convergence threshold for scf iterations
        diagonalization='david'                   !.Diagonalization
        electron_maxstep = 400                    !.Electron maximum step
/
        ATOMIC_SPECIES                    !.Type of atom used, atomic mass, and pseudopotential
        N   14.0067    N.pbe-n-kjpaw_psl.0.1.UPF
        Ca  40.078    Ca.pbe-spn-kjpaw_psl.0.2.3.UPF
        Na  22.98976928  Na.pbe-spn-kjpaw_psl.0.2.UPF

        ATOMIC_POSITIONS                                         !.Position of atom with respect to the crystal
        N  0.500   0.500   0.500
        N  0.000   0.000   0.000
        Ca  0.250   0.250   0.250
        Na  0.750   0.750   0.750

        K_POINTS (automatic)              !.Points in the reciprocal space used to sample the Brillouin zone
        8 8 8 0 0 0
```

Input File Description

**&control** Section - Specifies the type of calculation (SCF), file prefixes, and directories for pseudopotentials and temporary files.

This section contains parameters that control the general flow of the calculation.

**calculation='scf'**: Specifies that a self-consistent field (SCF) calculation will be performed. This means the program will iteratively solve the Kohn-Sham equations until the electron density converges.

**prefix='n2cana'**: Sets the prefix for output files, helping to identify files related to this specific calculation.

**pseudo_dir='/home/ishajemichael/PSEUDOPOTENTIALS/'**: Defines the directory where the pseudopotential files are located. Pseudopotentials approximate the effects of core electrons and simplify the calculations.

**outdir='./'**: Specifies the directory where temporary files will be stored during the calculation.

**&system** Section - Defines the Bravais lattice index (ibrav=2), lattice parameter (celldm(1)), number of atoms (nat) and types (ntyp) in the unit cell, electronic state occupations using Marzari-Vanderbilt smearing, and sets a plane-wave kinetic energy cutoff (ecutwfc) of 70 Ry.

This section defines the physical system to be studied, including lattice parameters, atom types, and electronic properties.

**ibrav=2**: Indicates the Bravais lattice type, with 2 representing a face-centered cubic (fcc) structure.

**celldm(1)=13.0684**: Sets the lattice parameter $a$ in atomic units (Bohr radii).

**nat=4**: Specifies the number of atoms in the unit cell.

**ntyp=3**: Indicates the number of different atomic species in the unit cell.

**occupations='smearing', smearing='mv', degauss=0.02**: Uses Marzari-Vanderbilt smearing for electronic state occupations with a smearing width of 0.02 Ry. This technique helps to achieve smooth convergence for metallic systems.

**ecutwfc=70**: Sets the plane-wave kinetic energy cutoff to 70 Ry. This parameter is crucial for the accuracy of the wavefunctions.

**&electrons** Section - Focuses on convergence settings for the SCF procedure, specifying the mixing factor (mixing_beta), diagonalization algorithm (Davidson), and the maximum number of iterations (electron_maxstep).

This section specifies the parameters for the SCF procedure.

**mixing_beta=0.7**: Sets the mixing factor for the charge density in the SCF iterations, aiding in the convergence of the electronic density.

**diagonalization='david'**: Chooses the Davidson diagonalization algorithm, which is efficient for large systems.

**electron_maxstep=400**: Limits the maximum number of SCF iterations to 400 to ensure the calculation converges properly.

**ATOMIC_SPECIES** Section - Details the types of atoms (N, Ca, Na), their masses, and corresponding pseudopotentials.

This section lists the atomic species present in the calculation along with their masses and the corresponding pseudopotential files.

**N 14.0067 N.pbe-n-kjpaw_psl.0.1.UPF**: Defines nitrogen with an atomic mass of 14.0067 amu, using the specified pseudopotential file.

**Ca 40.078 Ca.pbe-spn-kjpaw_psl.0.2.3.UPF**: Defines calcium with an atomic mass of 40.078 amu, using the specified pseudopotential file.

**Na 22.98976928 Na.pbe-spn-kjpaw_psl.0.2.UPF**: Defines sodium with an atomic mass of 22.98976928 amu, using the specified pseudopotential file.

**ATOMIC_POSITIONS** Section - Lists the positions of the atoms within the unit cell.

This section provides the positions of the atoms in the unit cell.

**N 0.500 0.500 0.500**

**N 0.000 0.000 0.000**

**Ca 0.250 0.250 0.250**

**Na 0.750 0.750 0.750**

These positions are given in crystal coordinates and define the specific arrangement of atoms within the FCC unit cell.

**K_POINTS** Section - Defines an 8x8x8 k-point grid for sampling the Brillouin zone, essential for accurately capturing the material's electronic properties.

This section specifies the k-point grid used to sample the Brillouin zone.

**8 8 8 0 0 0**: Defines an 8x8x8 Monkhorst-Pack grid with no shifts. This dense grid is crucial for accurately capturing the electronic properties of the material.

The input file carefully defines the parameters to perform a density functional theory (DFT) study on a half-metal alloy with a full Heusler structure of $N_2CaNa$. It covers fundamental aspects such as lattice structure, atomic positions, and electronic scavenging, providing SCF calculation settings for accurate modeling of electronic and magnetic material properties.

By specifying the lattice structure, the input file sets the Brave type and lattice parameters that define the geometry of the system. The positions of the atoms in the cell provide an accurate representation of the location of each element in the crystal structure, which is key to correctly modeling interatomic interactions and electronic structure. The use of the Marzari-Vanderbilt smearing ensures that the electron distribution is handled with appropriate accuracy, especially for metallic systems where it helps achieve better convergence.

In addition, the input file specifies important parameters such as the type of calculation, the cutoff energy of the wave function, the k-point grid for sampling the Brillouin zone, and the atom species and their corresponding pseudopotentials. These settings provide high precision in determining the electron density and energy levels, which is critical for understanding the material's physical properties. Thanks to such detailed adjustment, the SCF calculation allows obtaining reliable results, which are necessary for further research and practical application of the $N_2CaNa$ alloy in modern technologies.

## 3.0 Results

**3.1 Structural Properties:** The optimized lattice parameters, bond lengths, and angles of $N_2CaNa$ were determined

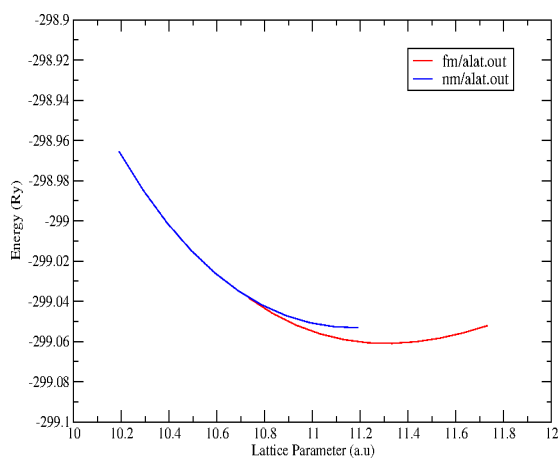 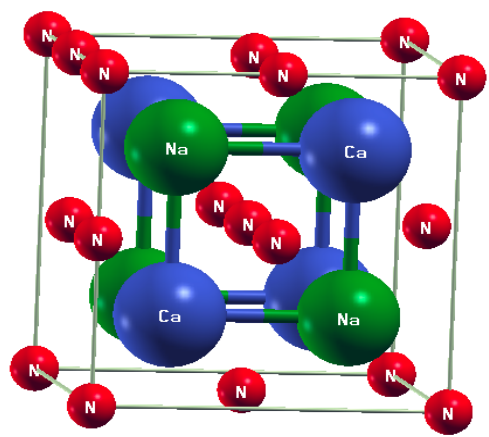

**1(A)**           **1(B)**

**Fig.1.** (A). Lattice parameter optimization curve for fm and nm $N_2CaNa$. (B). Crystal structure (ball and stick arrangement) of the Full-Heusler (FH) alloy $N_2CaNa$ viewed via XcrySDen.

**Table. 1.** Calculated Lattice Constant($a_o$), Bulk Modulus(B), and Pressure Derivative($B^/$) of Half-Heusler Alloy $N_2CaNa$

| Compounds | $a_o$(Å) | B(GPa) | $B^/$(GPa) | $E_g$(eV) |
|---|---|---|---|---|
| $N_2CaNa$ | 5.94376 | 61.6 | 4.95 | 29.90 |

We performed Series of self-consistent calculations to optimised the structure of $N_2CaNa$. The informational collections produced from the self-consistent energy computations were fitted to the third-order Birch-Murnaghan condition of state [3, 4] and the balance Lattice Constant($a_o$), Bulk Modulus(B), pressure

derivative( $B'$ ) and band-gap ($E_g$) were gotten. The mass modulus estimates how safe a material is to impressibility and its pressure derivative estimates its reaction to slight expansion in pressure. Fig. 2. (A & B), plot of energy and pressure against volume of $N_2CaNa$ respectively, shows that the higher the underlying volume of the material the more modest the energy in regard to the bulk and pressure derivative. From our determined qualities, we saw that $N_2CaNa$ will be effortlessly packed because of the small value of its bulk modulus.

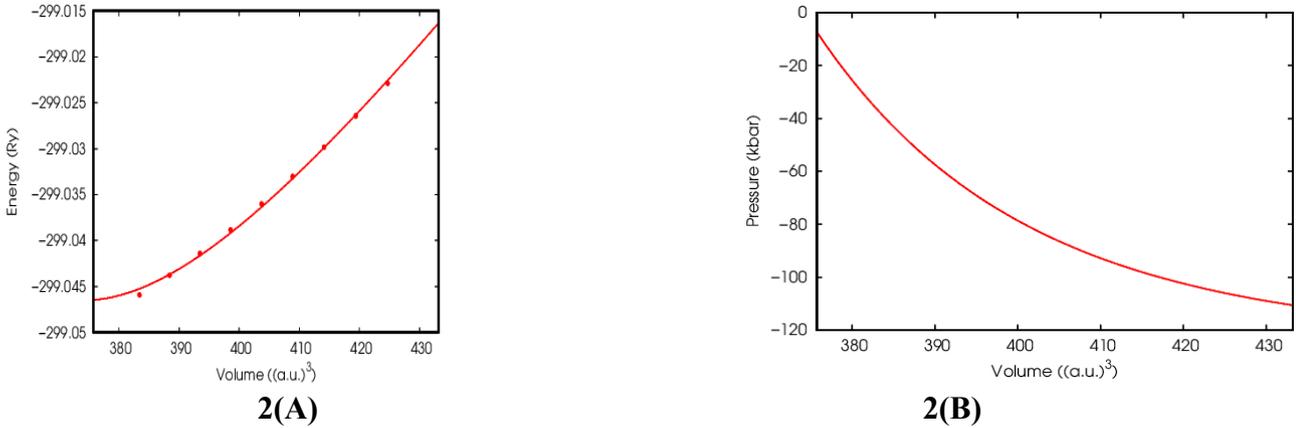

2(A)  2(B)

**Fig.2:** (A). Plot of Energy(RY) against Volume ((a.u)³) for $N_2CaNa$. (B). Plot of Presure(Kbar) against Volume ((a.u)³) for $N_2CaNa$

### 3.2 Density of State

Analysis of the electronic band structure revealed the full-metallic behavior of $N_2CaNa$ with a band gap in the minority spin channel and metallic behavior in the majority spin channel.

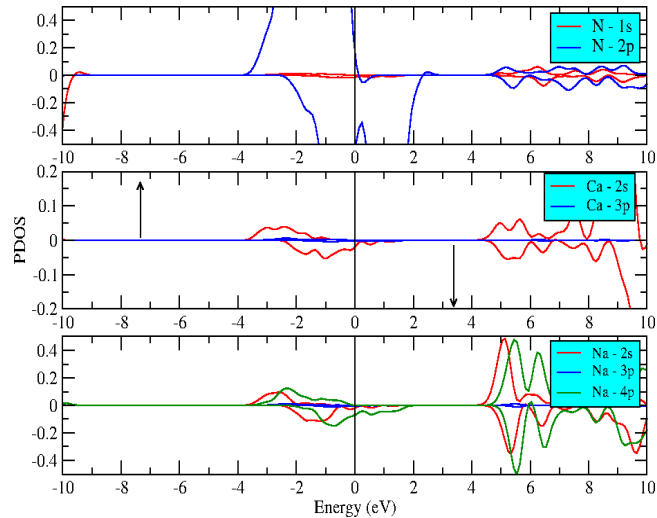

**Fig. 3:** Electronic density of states (DOS) of the element Nitrogen(N), Calcium(Ca), and Sodium(Na). They shows their P and s-orbital respectively.

The DOS investigate that N-2p, Ca-2s and Na-4p has more contributions. The brilion zone of the fermi-energy lines indicates the $N_2CaNa$ is half-metallic.

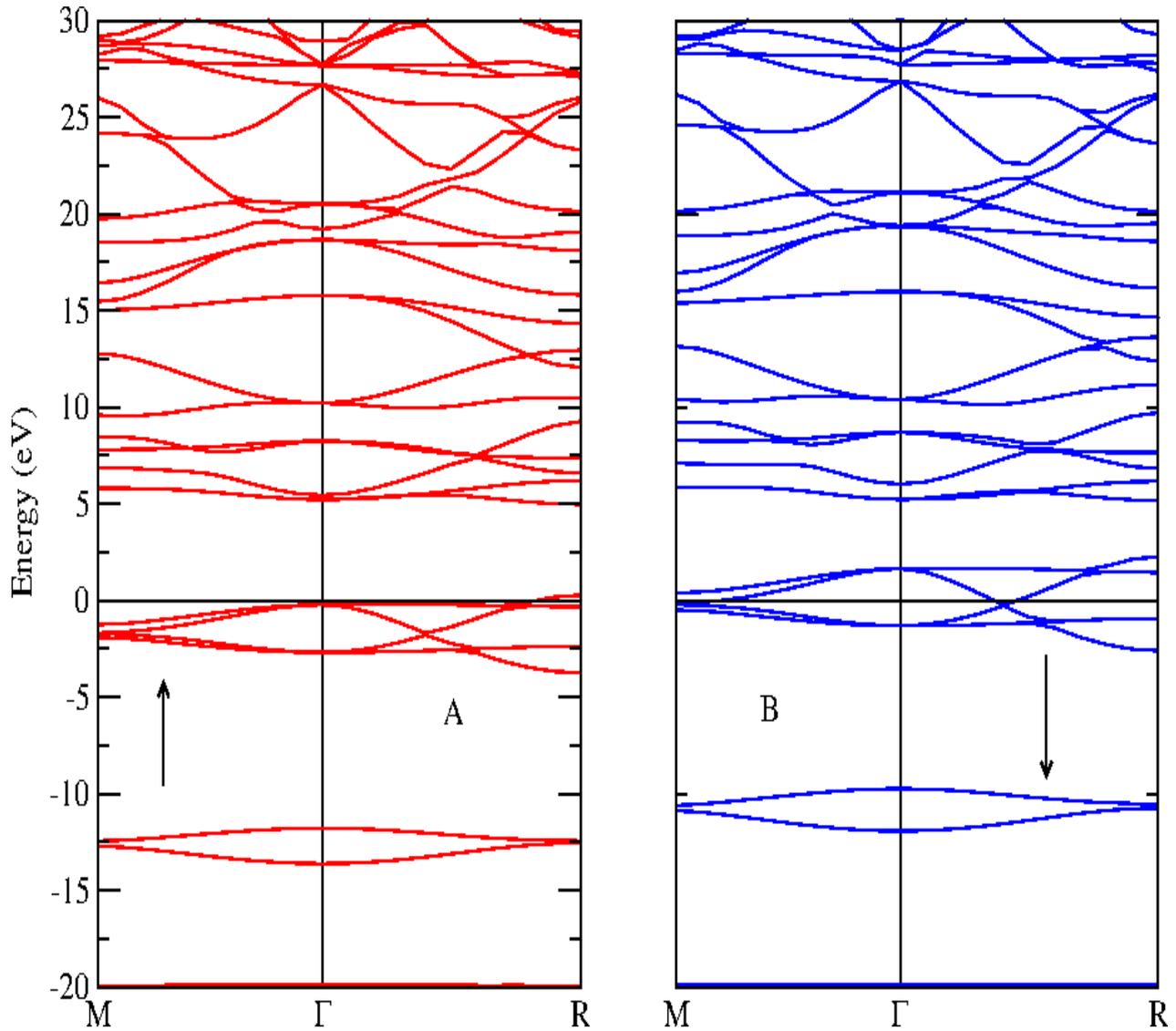

**Fig. 4.**(A). Spin up of the Electronic Band Structure. (B). Spin down of the electronic band structure. They shows that $N_2CaNa$ compound exhibit half-metallic properties with a narrow bandgap in the minority spin band and metallic behavior in the majority spin band.

The measured band gap of 5.08754eV was obtained for $N_2CaNa$ respectively. With the range of value, it is clear that both materials are narrow bandgap semiconductor full-Heusler alloys. From Fig. 3, we observed that the structures $N_2CaNa$ with their conduction band minimum (CBM) and the valence band maximum (VBM) is located at Gamma($\Gamma$)-point of the Brillouin zone. This indicates that the two alloys are direct bandgap semiconductors.

### 3.3 Mechanical Properties
The DFT calculations in this study revealed the following key findings: elastic constants, bulk modulus, and shear modulus were calculated to assess the mechanical stability of $N_2CaNa$**.** The bulk modulus and shear modulus values of N$_2$CaNa indicate the alloy's resistance to volume change and shear deformation, respectively. Young's modulus and Poisson's ratio values of N$_2$CaNa help characterize the material's mechanical response to different loads. We analysed the elastic properties of the materials in a bid to establish the mechanical stability of the compounds investigated. The Born stability criteria [10] is the generally accepted condition for mechanical stability for cubic structures, and it is given in Eqn. (1) as:

$$C_{11} > 0; C_{44} > 0; \frac{1}{2}(C_{11} - C_{12}) > 0; \frac{1}{3}(C_{11} + 2C_{12}) > 0 \tag{5}$$

the parameters $C_{11} + 2C_{12}$, $C_{11} - C_{12}$, and $C_{44}$ represents the non-degenerate, two-fold degenerate and three-fold degenerate levels of the bands in a cubic system. The three parameters correspond to the bulk, and the shear moduli of the crystal, and the cubic crystals become unstable if these values become negative [2]. It is, furthermore, required that $C_{12} < B < C_{11}$. Other elastic properties considered are bulk, shear, and young moduli. Computation of the moduli properties was performed using the following Equations (1), (2), (3), and (4).

**Table 2:** Calculated Elastic Constants ($C_{11}$, $C_{12}$ and $C_{44}$) and Zener Anisotropy factor(A) in Gpa for N$_2$CaNa Full-Heusler Alloy

| Compound | $C_{11}$ | $C_{12}$ | $C_{44}$ | A |
|---|---|---|---|---|
| N$_2$CaNa | 79.15 | 70.15 | 16.40 | 3.64 |

**Table 3:** Calculated Voigt Approximations: Shear Modulus(G), Young Modulus (E), Poisson's Ratio(n) and Bulk Modulus(B) in kbar for N$_2$CaNa Full-Heusler Alloy

| Compound | G | E | B | N | B/G | G/B |
|---|---|---|---|---|---|---|
| N$_2$CaNa | 12.71 | 35.64 | 60.59 | 0.415 | 4.766 | 0.209 |

**Table 4:** Calculated Reuss Approximations: Shear Modulus (G), Young Modulus (E), Poisson's Ratio(n) and Bulk Modulus(B) in kbar for N$_2$CaNa Full-Heusler Alloy

| Compound | G | E | B | n |
|---|---|---|---|---|
| N$_2$CaNa | 10.830 | 30.664 | 60.59 | 0.415 |

**Table 5:** Calculated Voigt-Reuss-Hill-Average of the two Approximations: Shear Modulus (G), Young Modulus (E), Poisson's Ratio(n) and Bulk Modulus(B) in kbar for N$_2$CaNa Full-Heusler Alloy

| Compound | G | E | B | n |
|---|---|---|---|---|
| N$_2$CaNa | 11.77 | 33.15 | 60.59 | 0.408 |

In [19] established 1.75 at the critical value from the ratio of bulk modulus to the shear modulus (B/G) of a material to be the reference value between brittleness and ductility in a material. When the B/G value is below 1.75 the material is brittle, above 1.74 the material is ductile. B/G ratio for $N_2CaNa$ is 4.766 as computed in Tab.2 above, hence the material is ductile. The brittleness or ductility can affect the failure state in fabrication processes.

### 3.4 Thermodynamic Properties

Heat capacity, entropy, and formation energy provided insights into the thermodynamic stability of $N_2CaNa$ under various conditions.

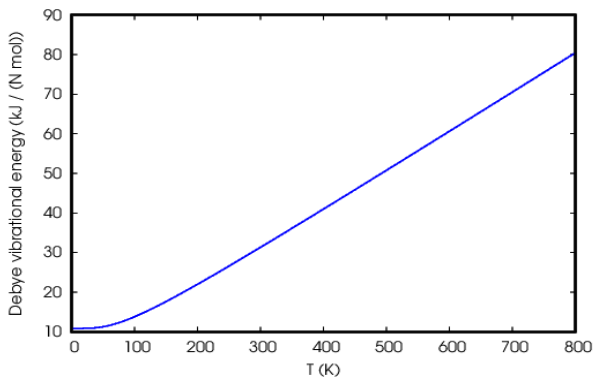

5(A)

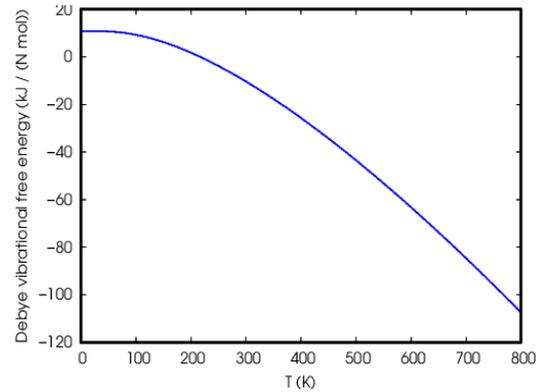

5(B)

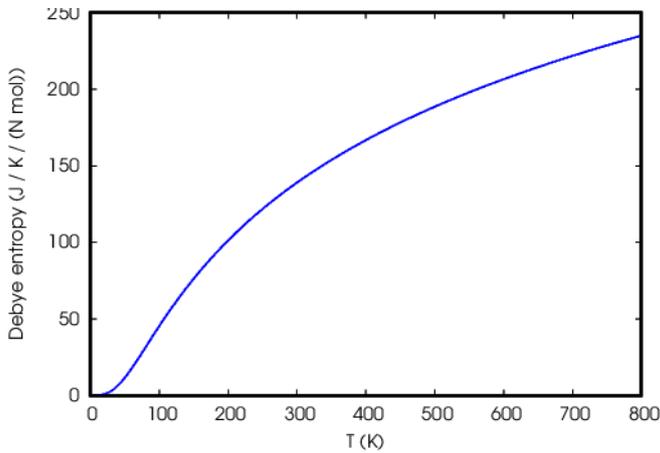

5(C)

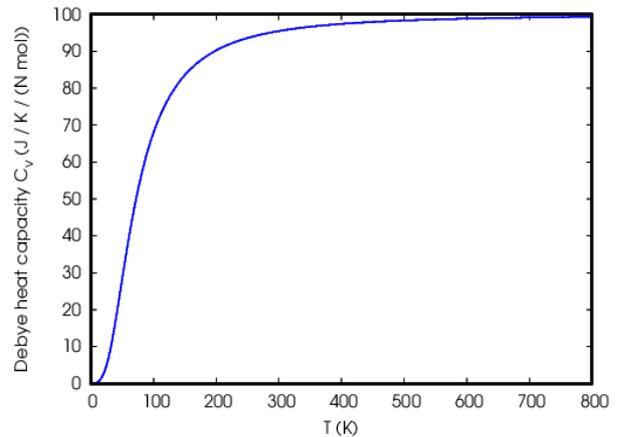

5(D)

**Fig. 5.** (A). Plot of Debye vibrational energy against temperature. (B). Plot of Debye vibrational free energy against temperature. (C). Plot of Debye entropy against temperature. Plot of Debye heat capacity against temperature.

Heat capacity, entropy, and formation energy provided insights into the thermodynamic stability of $N_2CaNa$ under various conditions. The thermodynamics properties results for $N_2CaNa$ is presented at Fig. 5(A to D) and discussion follows: Fig. 5(A) is a plot of Debye Vibrational energy at the 10 to 90 against temperature which shows that at low temperature the Debye vibrational energy is constant, but as the temperature increases rapidly the debye vibrational energy increases correspondingly. Furthermore, Fig. 5 (B) as the debye vibrational free energy increases the temperature decreases. Fig. 5(C) is similar to fig.5(A). While in Fig.5(D) the debye model correctly predicts the low temperature dependence of heat capacity, which is proportional to the debye $T^3$ law. Just like the Einstein model, it also recovers the Dulong-petit law at high temperatures. But due to simplifying assumptions, its accuracy suffers at intermediate temperatures. At room temperature debye specific heat ($C_v$) is 70($JK^{-1}N^{-1}mol$).

**4.0 Conclusion**

In conclusion, this DFT study provides comprehensive insights into the structural, electronic, mechanical, and thermodynamic properties of $N_2CaNa$. The results demonstrate its potential for applications in spintronics, structural engineering, and other fields requiring materials with tailored properties. Further experimental investigations are warranted to validate these theoretical findings and explore the practical applications of $N_2CaNa$.

The electronic properties, including band structure and density of states, highlight the potential of $N_2CaNa$ in spintronics. The material exhibits characteristics that are promising for spintronic devices, such as a suitable band gap and magnetic properties. These findings pave the way for developing next-generation electronic devices that leverage the unique electronic attributes of $N_2CaNa$.

The mechanical properties of $N_2CaNa$ have been thoroughly investigated, demonstrating high strength and elasticity. These attributes make it a potential candidate for applications requiring durable and resilient materials. The mechanical stability under different stress conditions further supports its applicability in structural engineering.

The study of thermodynamic properties, including specific heat and thermal expansion, suggests that $N_2CaNa$ can withstand varying temperature conditions while maintaining its integrity. This thermodynamic resilience is beneficial for applications in environments with fluctuating temperatures.

While this theoretical study [22-29] provides a solid foundation for understanding $N_2CaNa$, further experimental investigations are warranted to validate these findings. Experimental validation will not only confirm the theoretical predictions but also explore the practical feasibility and applications of $N_2CaNa$. Future research could focus on synthesizing the material and testing its properties in real-world scenarios.

Given the promising properties revealed by this study, $N_2CaNa$ has the potential to be a versatile material in several advanced fields. In spintronics, its electronic properties could lead to the development of more efficient and compact devices. In structural engineering, its mechanical robustness and thermodynamic stability could be leveraged to design more durable structures. The tailored properties of $N_2CaNa$ also open possibilities for other specialized applications, where specific material characteristics are required.

In summary, the DFT study of $N_2CaNa$ provides valuable insights that can guide future experimental research and practical applications. The combination of structural stability, electronic suitability for spintronics, mechanical strength, and thermodynamic resilience makes $N_2CaNa$ a material of significant interest.

**Acknowledgments**

Work was funded by the Department of Physics and Astronomy of the National Academy of Sciences of Ukraine under fundamental scientific program 0122U001501 (VS) and by US National Science Foundation (NSF) IMPRESS-U grant #2403609 via STCU project (KM).